\documentclass[12pt]{article}
\usepackage{amsmath,amsfonts,amssymb}
\usepackage{mathrsfs}
\usepackage{dsfont}
\usepackage[utf8x]{inputenc}
\usepackage{enumitem}
\usepackage[T1]{fontenc}
\usepackage{lmodern} \usepackage{ucs}
\usepackage{fullpage}
\usepackage{pifont}
\usepackage{graphicx}
\usepackage{caption}
\usepackage{subcaption}
\usepackage{float}

\usepackage{array}
\usepackage{multicol}
\usepackage{multirow}
\usepackage{color}
\usepackage{authblk}
\usepackage{esint}

\usepackage{tikz}
\usetikzlibrary{decorations}
\usetikzlibrary{decorations.pathreplacing,calc}
\usetikzlibrary{patterns}

\newtheorem{lemma}{Lemma}[section]

\newtheorem{proposition}[lemma]{Proposition}

\newtheorem{coro}[lemma]{Corollary}

\newcommand{\hslashslash}{%
  \raisebox{.9ex}{%
    \scalebox{.7}{%
      \rotatebox[origin=c]{18}{$-$}%
    }%
  }%
}

\def\dslash{%
  {%
   \vphantom{d}%
   \ooalign{\kern.05em\smash{\hslashslash}\hidewidth\cr$\mathrm d$\cr}%
   \kern.05em
  }}
  
\newcommand{\QED}{\mbox{}\hfill \raisebox{-0.2pt}{\rule{5.6pt}{6pt}\rule{0pt}{0pt}} \medskip\par}

\newcommand{\ds}{\displaystyle}
\newcommand{\ud}{\, {\mathrm{d}}}

\newcommand{\eps}{\epsilon}

\begin{document}

\title{Kinetic and macroscopic models for active particles exploring complex environments with an internal navigation control system}

\author{Luis G\'omez-Nava\thanks{ {\tt Luis\_alberto.Gomez\_nava@univ-cotedazur.fr}},
Thierry Goudon\thanks{ {\tt thierry.goudon@univ-cotedazur.fr}},
 Fernando Peruani\thanks{ {\tt Fernando.Peruani@univ-cotedazur.fr}}}

\affil{\small Universit\'e C\^ote d'Azur, Inria,  CNRS, LJAD,  

Parc Valrose, F-06108 Nice, France}

\date{}

\maketitle

\begin{abstract}
A large number of biological systems  -- from bacteria to sheep -- can be described as ensembles of self-propelled agents (active particles) with a complex internal dynamic that controls the agent's behavior: resting, moving slow, moving fast, feeding, etc.   
In this study, we assume that such a complex internal dynamic can be described by a Markov chain, which controls the moving direction, speed, and internal state of the agent. We refer to this Markov chain as the Navigation Control System (NCS). 
Furthermore, we model that agents sense the environment by considering that the  transition rates of the NCS  depend on local (scalar) measurement of the environment such as e.g. chemical concentrations, light intensity, or temperature.  
Here, we  investigate under which conditions the (asymptotic) behavior of the agents can be  reduced to an effective convection-diffusion equation for the density of the agents, providing effective expressions for the drift and diffusion terms.  
We apply the developed generic framework to a series of specific examples to show that in order to obtain a drift term two conditions should be fulfilled:  i) the NCS transition rates should depend on the agent's position, and ii) transition rates should be asymmetric. 
In addition, we indicate that the sign of the drift term -- i.e. whether agents develop a  positive or negative chemotactic response --  can be changed by modifying the asymmetry of the NCS or by swapping the speed associated to the internal states. 
The developed theoretical framework paves the way to model a large variety of biological systems and provides a solid proof that chemotactic responses can be developed, counterintuitively, by agents that cannot measure gradients and lack memory as to store past measurements of the environment. 
\end{abstract}

\section{Introduction}

Organisms, across scales, do not lock themselves in a behavioral task, but exhibit intermittent behavior: e.g. alternate between environment exploration, feeding, and resting~\cite{kramer2001behavioral}. 
This observation becomes apparent for large animals, but also applies to micro-organisms. 
Take as example bacteria as {\it E. coli}:  these bacteria swim, adhere to surfaces, and eventually form a biofilm or detach from one~\cite{ipina2019bacteria}. 
 Often, behaviors are strongly related, alternate each other over time, and complement each other to achieve a goal. 
 Environment exploration is a good example of this. 
 For example, sheep alternate short moving  and long stop phases as they forage, evaluating grass quality and eventually making a stop to feed~\cite{ginelli2015intermittent}. 
 In {\it E. coli}, surface exploration involves, contrary to sheep, long moving and short stop phases~\cite{ipina2019bacteria}. Stop phases are related to surface adhesion, 
  facilitate setting a new swimming direction, and arguably are used by the bacteria to test the surface properties. 
 Other bacteria as {\it P. putida} exhibit various swimming modes that involve, among other things, displacement modes at different speeds~\cite{hintsche2017polar}. 
 The mathematical modeling of intermittent motion, i.e. whether it is possible to conceive a generic theoretical framework to account such diversity of intermittent behaviors,  
 is unclear and represents a major theoretical challenge~\cite{nathan2008movement}. 
 The importance of such a generic theoretical framework would be paramount, providing a tool to describe biological system across scales, from  microorganisms to large vertebrates. 
 
 \begin{figure}[]
\centering
\includegraphics[width=0.8\textwidth]{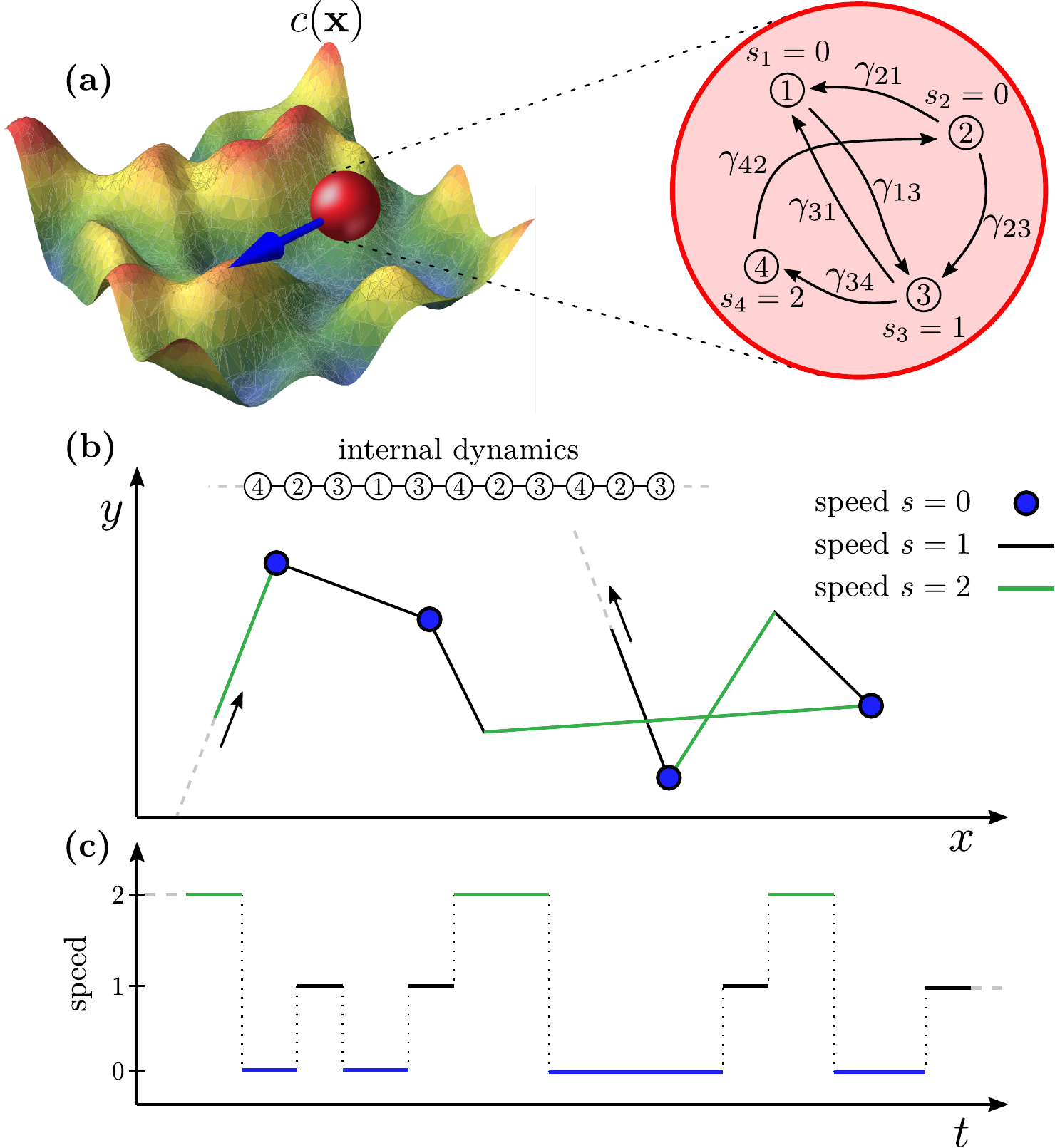}
\caption{(a) Scheme of one self-propelled particle navigating in a complex environment $c(\mathbf{x})$. An example of a possible navigation control system is highlighted in red. This internal system possesses four internal states, and six transition rates between them. Each of the states has a given value for the speed $s$ and the transition rates might involve also the change of direction of motion of the particle. 
(b) Example of a trajectory in a 2D space of a particle possessing the internal control system presented in (a). In this case, the internal dynamics is given as a sequence of states.
(c) Plot of the speed value of the particle as a function of time that results from the internal and the spatial dynamics presented in (b).}
\label{fig:Putida}
\end{figure}

Recently, it has been introduced a promising framework that describes moving biological entities as self-propelled agents -- a.k.a. active particles -- with  a complex internal dynamics 
that controls the agent's behavior~\cite{GNGP}. 
This complex internal dynamics is described by a Markov chain, which is also involved in the environment perception machinery of the agent 
by considering transition rates that depend on local, scalar, measurements of the environment. 
This framework has been proven successful to describe  intermittent motion in a variety of biological systems: {\it E. coli}~\cite{ipina2019bacteria}, {\it P. putida}~\cite{hintsche2017polar}, and sheep~\cite{GomezNava2020}. 
Here, we investigate the mathematical properties of an extended version of this framework. 
Assuming that the internal dynamics of agents is given by a Markov chain -- in the following referred to as Navigation Control System (NCS) -- that controls the moving direction, speed, and internal state of the agent, 
our main goal is to identify conditions that ensure that the  (asymptotic) behavior of the agents can be reduced to an effective convection-diffusion equation for the density  of the agents -- $(t,x)\mapsto \rho(t,x)$ -- of the form:
\begin{equation}\label{cd}
\partial_t\rho+\nabla_x\cdot(\rho U-D\nabla_x\rho)=0,
\end{equation}
We provide effective expressions for the drift  $U$ and diffusion $D$, which logically depend on the rates and network design of the NCS. 
By applying the  framework to a series of specific examples introduced in~\cite{GNGP}, we show that in order to observe $|U| \neq 0$ , i.e. chemotactic behavior, it is required that: 
i) a  number of internal states equal or larger than 2, ii) at least one transition rate that depends on the agent position (or local concentration of the external field), 
and iii) an asymmetric NCS structure (in a sense to be precise below).  
Furthermore, we indicate that the sign of $U$, which defines whether the chemotactic response is positive or negative, 
can be changed by modifying the asymmetry of NCS or by swapping the speed associated to the internal states. 

To appreciate the relevance of observing a non-vanishing drift term in this context, let us briefly review the assumption of previous chemotactic models. 
Let us recall that the here-considered agents cannot measure gradients directly; agents have access to local external concentrations only. Furthermore, 
we assume agents do not have the capacity to store past measurements of local concentrations -- as they move in space -- as to reconstruct gradients.  
And yet, we find that active agents with a NCS can respond to external gradients, exhibiting standard chemotactic responses. 
In the standard Keller-Segel (KS) model for chemotactic agents, it is assumed that agents instantaneously measure 
gradients and respond to it in such a way that  $U=\nabla_xc$, where $c$ is the chemo-attractant/repellent concentration~\cite{KS, Patlak}. 
And thus, observing a chemotactic response comes as no-surprise in KS model. 
The counterintuitive result in the KS model is that, if it is assumed that $c$ is produced by the individuals themselves  and $c$ defined through the Poisson equation $-\Delta_xc=\rho$,  
then concentration effects emerge depending on whether or not 
the initial mass exceeds a certain  threshold~\cite{Horstmann1, Horstmann2,BP}. 
The KS model can be derived following  a Boltzmann-like equation:
\begin{equation}\label{kin0}
\partial_t f+v\cdot\nabla_x f=Q(f)\, ,
\end{equation}
where $f(t,x,v)$ correspond to the distribution function of finding an individual at time $t$, at position $x$, moving in direction $v$, and 
$Q(f)$ is the reorientation operator defined by:
\begin{equation}\label{kin0}
Q(f)(t,x,v)=\ds\int k(x,v,v')f(t,x,v')\ud v'-f(t,x,v)\ds\int k(x,v',v)\ud v' .
\end{equation}
In order to obtain gradient sensing, the kernel
$k$ takes into account  memory effects, for instance 
by involving the concentration ahead $c(x+\eps v)$ and backward $c(x-\eps v')$, for some $\eps>0$, see~\cite{Chalub}. 
Other approaches include memory effects by assuming that   $k$ depend on the time derivative of the concentration along the pathways
of the individuals~\cite{BC, OH}. 
Other alternative is to assume that external measurements are stored by using an  additional variable $y$, with $y \in  \mathbb R$,
which leads to incorporate in  \eqref{kin0} a new term $\nabla_y\cdot(\mathscr G(y) f)$. In this way, the kernel $k$ depends on this variable $y$, see~\cite{PTV}. 
In summary, note that  in all these examples it has been assumed that either individuals can directly measure the chemical gradient or alternatively perform non-local measurements of the external concentration, or possess a memory kernel to store past concentration measurements.
Here, we will show that none of these assumptions is required for individuals to exhibit  chemotactic responses.
In summary, we  develop a generic theoretical framework to model  intermittent motion of individuals exploring complex environments and provide a novel, mathematical perspective on chemotaxis by proving agents with a NCS can display chemotactic behavior.
The paper is organized as follows. In Section \ref{sec:model}, we introduce the model and assumptions. In Section \ref{sec:rescaling} we present the main results obtained by coarse-graining the proposed microscopic model, leaving the derivations for later sections.   
In Section \ref{sec:applying} we apply these results to a series of illuminating examples to learn what agents with a NCS can do, and provide a summary of the obtained results in Section \ref{sec:conc}.  
All subsequent sections are devoted to the formal derivations of the results presented in Section \ref{sec:rescaling}: Section~\ref{S2} investigates the functional properties of the operator $Q$ that are needed to justify the analysis of the asymptotic regimes, while Section~\ref{S4} presents the proof of the asymptotic behavior, establishing the convergence towards the drift-diffusion equation.

\section{Active Particles with a Navigation Control System -- Model Description} \label{sec:model}

Our model consists of individuals that are characterized by a set of internal states that govern how they react to external signals.
The possible internal states associated to  the internal dynamics of the individuals are labelled by a discrete index $m\in \{1,...,M\}$, $M\in \mathbb N\setminus\{0\}$.
This set of internal states and the transition rates between them is going to be referred as the \textbf{N}avigation \textbf{C}ontrol \textbf{S}ystem of each particle.
The motion of the individuals is described by a ``velocity'' variable $v$, which ranges  a certain subdomain of $\mathbb R^N$, hereafter denoted $\mathscr V$, endowed with a suitable measure $\ud v$.
Of course, we can simply set $\mathscr V= \mathbb R^N$ and $\mathrm dv$ is the usual Lebesgue measure. 
But, as we shall see below, it can be relevant to consider situations where  $v$ lies in $\mathbb S^{N-1}$ -- in such a case it  is interpreted as the direction of motion of the particles -- or in a discrete subset of velocities in $\mathbb R^N$. 
In what follows, these frameworks are addressed in a unified fashion.
When the set  $\mathscr V$ is bounded it is particularly relevant to introduce 
an additional  parameter $s_m\geq 0$, indicating that the speed of the individuals 
can have a different magnitude depending on their current internal state.
For instance, certain states can be associated to significantly slower displacements, or even describe individuals at rest ($s_m=0$).
We consider the phase-space distribution of individuals: $P_m(t,x,v)$, for $m\in \{1,...,M\}$, $t\geq 0$, $x\in \mathbb R^N$, $v\in \mathscr V\subset \mathbb R^{N}$.
Thus, $\int_\Omega\int_{\mathscr O} P_m(t,x,v)\ud v\ud x$ gives the number of individuals in the state $m$ which, at time $t$, occupy a position $x\in \Omega\subset \mathbb R^N$, and move with the velocity $v\in \mathscr O\subset \mathscr V$.
The evolution of the population is driven by the PDE system
\begin{equation}\label{eq1}
\partial_t P_m+s_m v\cdot\nabla_x P_m=Q_m(P)
\end{equation}
where the interaction term describes both the mechanisms of change of direction and change of internal states.
In particular, it depends on all components of $P=(P_1,...,P_M)$.
Moreover, the rates of these modifications depend on the external signal, embodied into a scalar field $c:\mathbb R^N\rightarrow \mathbb R$.
In particular, the modeling assumes that the individuals are only sensitive to the local value of the signal, 
but they are not able to evaluate the gradient. 
To be more specific, we consider positively-valued functions
\[
(c,v,v')\in \mathbb R\times \mathscr V\times\mathscr V\longmapsto \gamma_{m,\ell}(c,v,v')>0
\] and
set 
\begin{equation}\label{defgam}\gamma_m(c,v)= \ds\sum_{\ell=1}^M\ds\int \gamma_{\ell, m}(c,v',v)\ud v'.\end{equation}
Next, we define
\begin{equation}
Q_m(P)(c,v)=\ds\sum_{\ell=1}^M\ds\int \gamma_{m,\ell}(c,v,v')P_\ell(v') \ud v' -\gamma_m(c,v)P_m(v).
\end{equation}
By virtue of \eqref{defgam}, the operator is mass-conservative in the sense that 
\[
\ds\sum_{m=1}^M\ds\int Q_m(P)(c,v)\ud v =0.\]
It means that this operator governs change of velocity and 
exchanges 
between the different sub-populations, ranked according to their internal state, but the total population satisfies a conservation property.
Consequently,
denoting 
\[\rho(t,x)=\ds\sum_{m=1}^M\ds\int P_m(t,x,v)\ud v,\qquad 
J(t,x)=\ds\sum_{m=1}^M\ds\int s_m v P_m(t,x,v)\ud v\] we get
\[\partial_ t \rho+\mathrm{div}_x J=0,
\] and, accordingly
\[
\ds\frac{\ud}{\ud t} \ds\sum_{m=1}^M\ds\iint
P_m(t,x,v)\ud v\ud x=0.\]
Relevant examples are thoroughly discussed below.

\section{Rescaling and macroscopic description: Drift and Diffusion Coefficients}\label{sec:rescaling}

The dynamics of the NCS -- i.e. the transitions from the internal states -- introduces a time scale associated 
to the internal dynamics of the agent. In the following, we assume that NCS characteristic time to be 
much shorter than the typical  time scale of the motion of the particle; moreover, we consider a large time scale of observation. 
These considerations lead to a rescaling of the equations, embodied into   a single scaling parameter $0<\eps\ll 1$, and the system can be rewritten:
\begin{equation}\label{eq1s}
\eps\partial_t P^\eps_m+s_mv\cdot\nabla_x P^\eps_m=\ds\frac1\eps Q_m(P^\eps).
\end{equation}
We are interested in the asymptotic behavior as $\eps$ goes to 0; we will establish that it can be described by a mere drift-diffusion equation for a macroscopic density $\rho$, with effective (drift and diffusion) coefficients depending on the rate coefficients $\gamma_{m,\ell}$. 
Namely, we shall see that 
$\rho^\eps=
\sum_{m=1}^M\int P^\eps_m\ud v$ converges (in a sense to be made precise) towards $\rho$, solution of the convection-diffusion 
\eqref{cd}.
The drift embodied into the effective macroscopic velocity field $x\mapsto U(x)$ is precisely due to the space dependence of the transition rates, that themselves depend on the signal $x\mapsto c(x)$. 
We shall see how relevant features, observable on the macroscopic scales, can be designed from the shape of the transition rates.

Let us briefly explain how the limit equation \eqref{cd} emerges in the regime $\eps\to 0$.
We expect that $Q_m(P^\eps)=\eps^2\partial_t P^\eps_m+\eps s_mv\cdot\nabla_x P^\eps_m$ tends to 0 as $\eps\to 0$.
Therefore the asymptotic dynamics is governed by the properties of the functions that make the interaction operator vanish.
Let us suppose that the $Q_m(P)$'s vanish iff the components $P_m$ are proportional to  certain functions $(x,v)\mapsto \mathscr E_m(x,v)$ satisfying
$$\mathscr E_m(x,v)>0,\qquad \ds\sum_{m=1}^M\int \mathscr E_m\ud v=1,\qquad Q_m(\mathscr E)=0.$$
This is the first key ingredient of the analysis.
Note that, because the interaction operator involves the space-dependent signal $x\mapsto c(x)$, the equilibrium $\mathscr E_m$ depends on the space variable.
We expand the solution of \eqref{eq1s}
as follows
$$P^\eps_m=P^{(0)}_m+\eps P^{(1)}_m+ \eps^2P^{(2)}_m+...
$$
We insert this expansion in \eqref{eq1s} and we identify terms arising with the same power of $\eps$.
At leading order, we get
$Q_m(P^{(0)})=0$, and we thus  infer $P^{(0)}_m(t,x,v)=\rho(t,x)\mathscr E_m(x,v)$.
Next, we get 
$$\begin{array}{lll}Q_m(P^{(1)})&=&s_m v\cdot\nabla_xP^{(0)}_m=s_m v\cdot\nabla_x (\rho\mathscr E_m)
\\
&=&\rho  s_m v\cdot\nabla_x \mathscr E_m + s_m v \mathscr E_m\cdot\nabla_x\rho.
\end{array}
$$
That the equilibrium function $\mathscr E_m$ depends on the space variable is the source of the drift term.
Indeed, the second key ingredient of the analysis  is the possibility to invert the equations $Q_m(P)=R_m$, provided the compatibility condition $\sum_{m=1}^M\int s_mvR_m \ud v=0$ holds.
Therefore, the equilibrium is requested to fulfil this condition 
$$\sum_{m=1}^M\int s_m v\mathscr E_m \ud v=0$$
so that we can find the corrector $P^{(1)}_m(t,x,v)=\chi_m(x,v)\cdot \nabla_x\rho(t,x)+ \lambda_m (x,v)\rho(t,x)$
where 
$Q_m(\chi)=s_m v \mathscr E_m$, and $Q_m(\lambda)=s_m v \cdot\nabla_x\mathscr E_m$.
Finally, the mass balance principle applied to $Q_m(P^{(2)})=\partial_t P^{(0)}_m +v\cdot\nabla_x P^{(1)}_m$
yields \eqref{cd}
with the following expression of the diffusion and  transport coefficients
\begin{equation}\label{eq:DiffCoeff}
D(x)=-\ds\sum_{m=1}^M \ds\int s_m  v\otimes 
\chi_m(x,v) \ud v
\end{equation}
and
\begin{equation}\label{eq:DiffConv}
U(x)=\ds\sum_{m=1}^M \ds\int  s_m v\lambda _m(x,v) \ud v.
\end{equation}
This can be understood by considering the mass conservation relation as well.
Indeed, on the one hand
$\rho^\eps(t,x)=\sum_{m=1}^M\int P^\eps_m(t,x,v)\ud v$ and 
$J^\eps(t,x)=\frac1\eps \sum_{m=1}^M\int  s_m vP^\eps_m(t,x,v)\ud v$ satisfy
$$\partial_t\rho^\eps+\nabla_x\cdot J^\eps=0.$$ On the other hand,
we guess that $$P^{\eps}_m(t,x,v)=\rho^\eps(t,x) \mathscr E_m(x,v) + \eps G^\eps_m(t,x,v)$$
 and \eqref{eq1s} casts as 
 $$
\eps\partial_t P^\eps_m+s_mv\cdot\nabla_x P^\eps_m= Q_m(G^\eps).
$$
Assuming that all quantities admit limits, we obtain 
$Q_m(G)=s_mv\cdot\nabla_x(\rho \mathscr E_m)$,
the solution of which can be identified as above.
The knowledge of the remainder  $G$ allows us  to conclude by passing to the limit in the mass conservation equation since, owing to to the compatibility condition, we get
$$J^\eps = \ds\sum_{m=1}^M\ds \int  s_m v G^\eps_m\ud v\xrightarrow [\eps\to 0]{}
\ds\sum_{m=1}^M\ds \int  s_m v G_m\ud v
=\rho U -D\nabla_x \rho.$$
We point out again that the drift only comes from the space dependence of the equilibrium, induced by the fact the transition rates depend on the signal, 
but we did not need any scaling of the signal, nor memory effect. 
We stress that transition rates dependent on the signal, and consequently on space, is not a sufficient condition to obtain a non-vanishing drift term. 
More than one internal state is required to induce such a drift, and in addition, transition rates have to be asymmetric. 
This is illustrated with a series of simple illuminating examples in the next sections. 
A rigorous justification of the asymptotic regime summarized in this section is presented in Section~\ref{S2} and~\ref{S4}.

\section{Applying the formalism to key examples: Learning what agents with a NCS can do}
\label{sec:applying}

Here, we use the developed formalism to study two NCS designs that operate with two states initially proposed in~\cite{GNGP}.   
For simplicity, we  focus on one-dimensional spatial systems.
Extensions for larger number of states of higher dimensions are straightforward. 

\begin{figure}[t]
\centering
\includegraphics[width=1\textwidth]{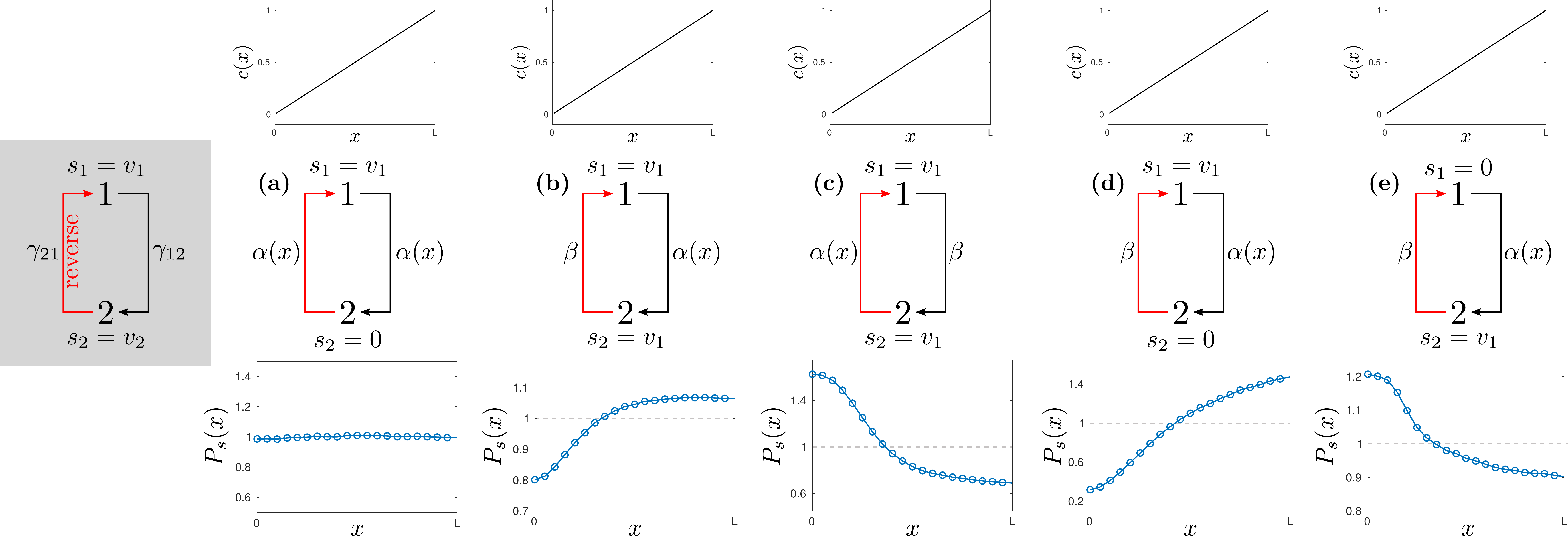}
\caption{The simplest non-trivial NCS with two internal states and two transitions rates. 
The grey box provides a general sketch of this two-state NCS. 
The transition from state 2 to 1 -- colored in red -- triggers (in all examples) a reversal of the direction of motion of the particles. 
Panels (a) - (e) show that agents exposed always to the same external field $c(x)$ (see upper panel row) respond differently to the external signal depending 
on the design of the NCS: the stationary distribution $P_s(x)$ is provided for all examples (see lower panel row). 
Note that if transition rates are identical, there is no chemotactic response, while for asymmetric transition rates -- i.e. $\alpha \neq \beta$ -- agents display chemotactic responses. 
Interestingly, by exchanging $\alpha$ and $\beta$ (maintaining always the fact that the transition $2 \to 1$ triggers a reversal) or by exchanging the speed associated to state 1 and 2, it is possible to switch from a positive to a negative chemotactic response. 
The Individual Based Model (IBM) simulations were performed using parameters $N = 10000$ particles, $v_1 = 0.03$, $L = 1$, $\alpha(x) = x/L$, $\beta = 0.1$ and an observation time of $T_{\text{obs}} = 10000$.} 
\label{fig:motif1}
\end{figure}

\subsection{NCS design 1: non-adaptive chemotactic behaviors}

We assume that NCS possesses two states, $1$ and $2$, with associated speeds $s_1$ and $s_2$, respectively.  
There are only two possible transitions: one from state $1$ and $2$, characterized by rate  $\gamma_{12}$, 
and another one from  $2$ and $1$ with rate $\gamma_{21}$. 
The latter transition triggers a reversal of the direction of the active motion, see Fig.~\ref{fig:motif1}. 
To ease the notation and avoid using sub-indices, we define $\gamma_{12}=\alpha$ and $\gamma_{21}=\beta$. 
Recall that transition rates in general depend on spatial position of the agent $x$ (through the local value of the external concentration). 
On a spatial one-dimensional space, we can distinguish left moving and right moving particle:  $\ud v=\frac12(\delta(v=-1)+ \delta(v=+1))$. 
Then, we can use $P^{+}_1$ and $P^{-}_1$ to denote right and left moving particles, respectively, in state $1$, and 
$P^{+}_2$ and $P^{-}_2$ for right and left moving particles in state $2$. 
Using this definitions, the system dynamics is given by the following set of PDEs: 
\begin{equation}\label{eq:motif1}
\begin{array}{l}
\partial_t \begin{pmatrix}P_1^+ \\ P_1^- \\ P_2^+ \\ P_2^-\end{pmatrix}
+\Lambda
 \partial_x \begin{pmatrix}P_1^+ \\ P_1^- \\ P_2^+ \\ P_2^-\end{pmatrix}
 = 
\begin{pmatrix}  -\alpha &  0 &  0 & \beta  \\ 0 & -\alpha & \beta &  0   \\ \alpha&  0&  -\beta & 0  \\ 0 & \alpha &   0&  -\beta\end{pmatrix}
 \begin{pmatrix}P_1^+ \\ P_1^- \\ P_2^+ \\ P_2^-\end{pmatrix},
\end{array}
\end{equation}
with $\Lambda=\mathrm{diag}(s_1,-s_1,s_2,-s_2)$.

\subsection{Rescaling and effective coefficients for NCS design 1}

We rescale Eq.~(\ref{eq:motif1}) as indicated by expression (\ref{eq1s}):  
\begin{equation}\label{2v_res}
 \eps\partial_t P^\eps+\Lambda\partial_x P^\eps=\ds\frac1\eps Q P^\eps.
\end{equation}
 We verify that: 
 \[\mathrm{Ker}(Q)=\mathrm{Span}\{\mathscr E\},\qquad \mathscr E=\ds\frac{1}{2(\alpha+\beta)}\begin{pmatrix}\beta\\ \beta\\ \alpha\\ \alpha\end{pmatrix},\]
 and 
 \[\mathrm{Ran}(Q)=\big\{R\in \mathbb R^4,\ R_1^++R_1^-+R_2^++R_2^-=0\big\}.\]
 Let $$
 \begin{array}{l}
 \rho^\eps(t,x)=P^{+,\eps}_1+P^{-,\eps}_1+ P^{+,\eps}_2+P_2^{-,\eps},\\[.3cm]
 J^\eps(t,x)=\ds\frac1\eps(s_1P^{+,\eps}_1-s_1 P^{-,\eps}_1+ s_2P^{+,\eps}_2-s_2P_2^{-,\eps}),
 \end{array}
 $$
that satisfy the conservation law
\[\partial_t \rho^\eps+\partial_x J^\eps=0.\]
We can rewrite \eqref{2v_res} as 
\begin{equation}\label{2v_res2}\eps\partial_t P^\eps+\Lambda\partial_x P^\eps= Q G^\eps
\end{equation}
where 
\[G^\eps=\ds\frac1\eps(P^\eps-\rho^\eps \mathscr E)\]
can be expected to remain bounded.
Passing formally to the limit in  \eqref{2v_res2}, we are led to 
\[\Lambda\partial_x (\rho\mathscr E)= Q G=\Lambda\mathscr E\partial_x \rho + \rho\Lambda\partial_x \mathscr E.\]
Observe that the sum of the components of $\Lambda\mathscr E$ vanishes, so that it lies in $\mathrm{Ran}(Q)$ and we can find vector valued quantities
$\chi$ and $ \lambda$ such that 
$$Q\chi=\Lambda\mathscr E,\qquad Q\lambda=\Lambda\partial_x \mathscr E.$$

For the computations, we can take advantage of the fact that the subspace generated by $(1,-1,0,0)$ and $(0,0,1,-1)$ is stable by the action of the matrix $Q$.
To be specific, we simply find
\[\chi= \ds\frac{1}{2(\alpha+\beta)}
\Big(\ds\frac{s_2\alpha-s_1\beta}{2\alpha},\ \ds\frac{s_1\beta-s_2\alpha}{2\alpha},\ -\ds\frac{s_2\alpha+s_1\beta}{2\beta},\ \ds\frac{s_2\alpha+s_1\beta}{2\beta}\Big),
\]
and
\[\begin{array}{lll}
\lambda&=& \ds\frac{1}{4(\alpha+\beta)}
\Big(\ds\frac{s_2\partial_x\alpha-s_1\partial_x\beta}{\alpha},\ -\ds\frac{s_2\partial_x\alpha-s_1\partial_x\beta}{\alpha},\
 -\ds\frac{s_2\partial_x\alpha+s_1\partial_x\beta}{\beta},\ \ds\frac{s_2\partial_x\alpha+s_1\partial_x\beta}{\beta}\Big)
\\
&&
- \ds\frac{\partial_x\alpha+\partial_x\beta}{4(\alpha+\beta)^2}\Big(\ds\frac{s_2\alpha-s_1\beta}{\alpha},\
-\ds\frac{s_2\alpha-s_1\beta}{\alpha},\ -\ds\frac{	s_2\alpha+s_1\beta}{\beta},\ \ds\frac{s_2\alpha+s_1\beta}{\beta}
\Big).
\end{array}
\]
Thus, we arrive at $G=\chi\partial_x\rho+\rho\lambda$.

Accordingly, we also have
\[J^\eps=s_1G_1^{+,\eps}-s_1G_1^{-,\eps}+s_2G_2^{+,\eps}-s_2G_2^{-,\eps},\]
which tends to 
$$-D\partial_x\rho+ \rho U$$
with
\begin{eqnarray}
\begin{split}
U &= (s_1\lambda_1^+-s_1\lambda_1^-+s_2\lambda_2^+-s_2\lambda_2^-)\\
    &= \ds\frac{s_2(s_1\beta-s_2\alpha)\partial_x\alpha-s_1(s_2\alpha+s_1\beta)\partial_x\beta}{2(\alpha+\beta)\alpha\beta}
+ \ds\frac{s_2^2\alpha^2+s_1^2\beta^2}{2(\alpha+\beta)^2\alpha\beta}(\partial_x\alpha+\partial_x\beta),
\end{split}
\label{eqn:drift_expression}
\end{eqnarray}
and
\begin{equation}
D=-(s_1\chi_1^+-s_1\chi_1^-+s_2\chi_2^+-s_2\chi_2^-)=\ds\frac{1}{2(\alpha+\beta)}\Big(s_1^2\ds\frac\beta\alpha+s_2^2\ds\frac{\alpha}{\beta}\Big)
=\ds\frac{s_2^2\alpha^2+s_1^2\beta^2}{2(\alpha+\beta)\alpha\beta}.
\label{eqn:Diff_coeff}
\end{equation}
We are thus led to the  drift-diffusion  equation \eqref{cd}, which  describes the asymptotic behavior of the system.  
We discuss the obtained results below, after introducing and studying NCS design 2.

\subsection{NCS design 2: adaptive chemotactic behaviors}\label{subsec:adapt}

Particles operating with NCS design 1 respond to the an external concentration via $U(x)$ as well as via $D(x)$. 
Here, we will introduce the concepts of chemotaxis and chemokinesis. 
Assume $U(x)=0$ and $D(x) \neq 0$ such that  
particles certainly respond to the external concentration via $D(x)$. 
Such a behavior is usually referred to as chemokinesis. Usually $D(x)$ depends on the value of the external concentration as occurs with NCS design 1: 
$D(x)$ is function of $\alpha$ and $\beta$, which in turn depend on the external field $c(x)$. 
Due to this dependency on $c(x)$, it is said that the behavior is non-adaptive. 
On the other hand, if $D(x) = D_0$, with $D_0$ a constant, and $U(x) \neq 0$ and such that $U \propto \partial_x c$, 
then the response to the external concentration is via (pure) chemotaxis. 
And since particles respond mainly to $\partial_x c$, and only weakly to $c$, it is said that the behavior is adaptive: 
the background level of the external concentration $c$ does not strongly affect the behavior of the agents that respond mainly to $\partial_x c$.  
In the following, we show that it is possible to obtain adaptive chemotactic behavior using a NCS. 
Again, we use two states, $1$ and $2$. For simplicity, we assume that $s_1=s_2=1$. 
The fundamental difference with NCS 1 is that here we consider 
three possible transitions: one transition from state $1$ to $2$, characterized by rate  $\alpha$, 
and two transitions from $2$ to $1$, characterized by rates $\beta$ and $\gamma$. 
The reversal in the moving direction is triggered only by one of the two transitions $2\to 1$, the one associated to transition rate $\beta$. 
The resulting system in a one-dimensional space is given by:
\begin{equation}\label{eq:adaptive}
\begin{array}{l}
\partial_t \begin{pmatrix}P_1^+ \\ P_1^- \\ P_2^+ \\ P_2^-\end{pmatrix}
+\Lambda
 \partial_x \begin{pmatrix}P_1^+ \\ P_1^- \\ P_2^+ \\ P_2^-\end{pmatrix}
 = 
\begin{pmatrix}  -\alpha &  0 &  \gamma & \beta  \\ 0 & -\alpha & \beta &  \gamma   \\ \alpha&  0&  -\beta-\gamma & 0  \\ 0 & \alpha &   0&  -\beta-\gamma\end{pmatrix}
 \begin{pmatrix}P_1^+ \\ P_1^- \\ P_2^+ \\ P_2^-\end{pmatrix} \,,
\end{array}
\end{equation}
with $\Lambda=\mathrm{diag}(1,-1,1,-1)$. 
The third transition associated to $\gamma$ by itself cannot ensure adaptive chemotactic behavior. In addition, conditions on $\alpha$, $\beta$, and $\gamma$ have to be requested. 
We discuss these conditions after rescaling NCS 2 for three generic transition rates. 

\subsection{Rescaling  and effective coefficients for NCS design 2}

We rescale Eq.~(\ref{eq:adaptive}) as indicated by expression (\ref{eq1s}) and verify that: 
\[\mathrm{Ker}(Q)=\mathrm{Span}\{\mathscr E\},\qquad \mathscr E=\ds\frac{1}{2(\alpha+\beta+\gamma)}\begin{pmatrix}\beta+\gamma\\ \beta+\gamma\\ \alpha\\ \alpha\end{pmatrix},\]
 and
 \[\mathrm{Ran}(Q)=\big\{R\in \mathbb R^4,\ R_1^++R_1^-+R_2^++R_2^-=0\big\}.\]
With similar notations as above, we get 
\[\chi= \ds\frac{1}{2(\alpha+\beta+\gamma)}(X,-X,Y,-Y),
\quad X = \ds\frac{\beta-\gamma}{2\beta}-\ds\frac{(\gamma+\beta)^2}{2\alpha\beta}
,\quad
Y=-\ds\frac{\alpha+\beta+\gamma}{2\beta}.\]
Therefore, we are led to the following definition for the effective diffusion and convection coefficients:
\[\begin{array}{l}
D= \ds\frac{\alpha^2+2\alpha\gamma+(\gamma+\beta)^2}{2\alpha\beta(\alpha+\beta+\gamma)},\\[.3cm]
U=\ds\frac12\partial_x\Big(\ds\frac{\alpha+\gamma}{\alpha\beta}\Big) - \ds\frac{\beta+\gamma-\alpha}{2\alpha^2(\alpha+\beta+\gamma)}
\partial_x \alpha+\partial_x D.
\end{array}\]
As mentioned above, it is possible to request conditions on the rate $\alpha$, $\beta$, and $\gamma$ 
in order to obtained an adaptive chemotactic behaviors. 
Let us recall that $\alpha$, $\beta$, and $\gamma$  depend on the external concentration field $c$. 
Then, given a function $c\mapsto \beta(c)$ we request $\beta(c) \in [\beta_\star,\beta^\star]$ and set $\alpha(c)=\frac{\beta_\star\beta^\star}{\beta(c)} $ and $\gamma(c)=2\beta_\star
-\alpha(c)-\beta(c)$ (which remains non negative).
Then, the diffusion becomes $D=\frac12\frac{\beta^{\star 2}}{\beta_\star\beta^\star(\beta_\star+\beta^\star)}$, which does not depend on $c$, while we get a space-dependent convection $U(x)=-\frac1{\beta_\star+\beta^\star}\frac{\beta'(c(x))}{\beta(c(x))}\partial_x c(x)$, which creates an adaptive dynamics controlled by the gradient of $c$.

\section{Concluding remarks on the behavior of agents with NCS}
\label{sec:conc}
 
The developed theoretical framework allows us to analyze active agents with an arbitrary number of internal states interconnected and controlled by the NCS.    
In particular, we determined conditions that ensure that the asymptotic behavior of the agents can be reduced to a convection-diffusion equation, providing 
expressions for  the drift $U$ and diffusion coefficient $D$. 
By applying these results to specific examples, we obtained a series important remarks on the asymptotic behavior to this type of agents. 

\begin{itemize}
\item [{\bf{ (R1)}}] {\it At least two internal states are required to obtain a non-vanishing drift term $U$}. \\
If we assume only one internal state, the NCS degenerates to $\begin{pmatrix}-\alpha & \alpha\\ \alpha & -\alpha\end{pmatrix}$ and even if the coefficient $\alpha$ depends on $c(x)$, the equilibrium $\mathscr E=(1,1)$ does not. 
And thus, we cannot observe a macroscopic drift $U$ induced by the external concentration $c$.
\item [{\bf{(R2)}}] {\it Non-symmetric transition rates are required to obtain a non-vanishing drift term $U$}. \\
Assuming two internal states and transition rates  $\alpha$ and $\beta$ that depend on the external concentration $c$, 
we have shown that for  $\alpha=\beta$, $U$ vanishes, even when $s_2\neq s_1$.  
To observe a non-vanishing drift, it is thus necessary to assume  $\alpha\neq \beta$, and additionally request that at least one of the transition rate is $c$-dependent. 
See Fig.~\ref{fig:motif1}.

\item [{\bf{(R3)}}] {\it The sign of $U$ is controlled by the design of the NCS and the speed  values associated to the internal states}. \\
By either exchanging $\alpha$ and $\beta$ or by exchanging the speed associated to speed 1 and 2, it is possible to invert the sign of $U$.  
This highlights the importance of the design of the NCS in the chemotactic response and indicates that positive and negative chemotactic responses (i.e. up-gradient or down-gradient biased motion)  
can be induced by altering the NCS design: exchanging of rates or state speeds can invert the chemotactic behavior. 

\item [{\bf {(R4)}}] {\it The asymptotic spatial distribution of agents}.\\ 
By setting $P_s(x)=\exp\Big(\int_0^x \frac{U}{D}(y)\ud y\Big)$, we can rewrite the limit equation as \[\partial_t\rho - \partial_x \Big(DP_s\partial_x \Big(\ds\frac\rho P_s\Big)\Big)=0,\] which allows us to identify the equilibrium spatial distribution of the agents.

\item [{\bf{(R5)}}] {\it Adaptive chemotactic responses}. \\
We have shown in subsection \ref{subsec:adapt} that by introducing three transition rates into the two-state NCS, 
it is possible to conceive transition rates that lead to adaptive chemotactic responses. 
This implies that agents operating by  two-state NCS can exhibit the same chemotactic performance independently of the 
background level of $c(x)$, exhibiting a drift that is proportional to the external field gradient, i.e. $U \propto \partial_x c$. 
\end{itemize}

In summary, our study provides a solid mathematical understanding of the asymptotic behavior of agents operating by a NCS, providing a generic framework to model and understand intermittent collective motion in biological systems, and a novel, mathematical perspective on chemotaxis, 
by showing that neither memory (in the sense of storage of past measurements) nor non-local external field measurements (to directly evaluate gradients) are required to observe such type of behaviors.

\section{Study of the interaction operator}
\label{S2} 

This Section is devoted to the analysis of the interaction operator $Q$.
For future purposes, it is important to bear in  mind that the coefficients of the operator depend on the space variable $x$.
To make the notation less cluttered, in this Section we do not mention this parameter,
assuming implicitely that all estimates discussed below hold uniformly with respect to $x$.

\subsection{Equilibrium and dissipation}

We assume the existence of an equilibrium
\begin{equation}\tag{{\bf {A1}}}\label{A1}\left\{\begin{array}{l}
\textrm{There exists a $M$-uplet of functions $\mathscr E_m:\mathbb R\times \mathbb R^N\rightarrow (0,\infty)$ such that }
\\
\gamma_m(v)\mathscr E_m(v)=\ds\sum_{\ell=1}^M \ds\int \gamma_{m,\ell}(v,v')\mathscr E_\ell(v')\ud v',
\\
\ds\sum_{m=1}^M\ds\int \mathscr E_m(v)\ud v=1,
\\
\ds\sum_{m=1}^M\ds\int \Big(\gamma_m(v)+\ds\frac{1}{\gamma_m(v)}\Big)\mathscr E_m(v)\ud v=\mu \textrm{ is finite}.
\end{array}\right.
\end{equation}
This property can be checked depending on the coefficients $\gamma_{m,\ell}$.
For instance, it holds assuming that the $\gamma_{m,\ell}'s$ are continuous and positive, with $V=\mathbb S^{N-1}$.
More generally, it suffices to check that 
\[
\ds\sum_{m=1}^M \ds\int 
\ds\sup_{v'\in \mathscr V, \ell} \left(
\ds\frac{1+\gamma_m(v)}{1+\gamma_m(v')}\ds\frac{\gamma_{m,\ell}(v,v')}{\gamma_m(v)}
\right)\ud v<\infty.\]
It permits us to apply the Krein-Rutman theorem, see \cite[Theorem V.6.6]{Sch}, since a power of the underlying linear operator 
(which is positive)
is compact, see \cite[Chapter~11, \S 2, Ex.~E]{za}. We also refer the reader to \cite{Ev} for further characterization of the compactness of integral operators.
 In order to obtain useful dissipation estimates, we shall also need the following strengthened assumption:
\begin{equation}\tag{{\bf {A2}}}\label{A2}\left\{\begin{array}{l}
\hbox{\rm 
There exists a positive constant $\kappa$ such that}
\\
\displaystyle \mathscr E_m(v) \leq \kappa \, 
\left(\gamma_m(v')+\frac{1}{\gamma_m(v')}\right)\, \frac{1}{\gamma_m(v)}\, \gamma_{m,\ell}(v,v').
\end{array}
\right.
\end{equation}

Let us collect here all the technical 
assumptions that will be necessary to justify the derivation of a macroscopic model.
We assume that
 there exists two positive constants $\mu_1$ and $\mu_2$ such that
\begin{equation}\tag{{\bf {B1}}}\label{B1} s_m |v| \, |\nabla_x \mathscr E_m(x,v) | \leq \mu_1 \gamma_m(x,v) \, \mathscr E_m(x,v), 
 \hbox{\rm a.e.},
\end{equation}
\begin{equation}\tag{{\bf {B2}}}\label{B2}\qquad
s_m^2
 \int_{\mathscr V} |v|^2 \frac{\mathscr E_m(x,v)}{\gamma_m(x,v)} \ud v \leq \mu_2,
\quad   \hbox{\rm a.e.},
\end{equation}
\begin{equation}\tag{{\bf {B3}}}\label{B3} \ds\sum_{m=1}^M \ds\int_{\mathscr V} s_m v\, \mathscr E_m(x,v)\, \ud v=0,\quad    \hbox{\rm a.e.}
\end{equation}
We point out that $v F(x,v)$ is integrable for $x$ a.e. because of \eqref{A2} and \eqref{B2}, so that
\eqref{B3} makes sense.
Finally, we need a geometrical assumption on the set of velocities
\cite{DGP, LiTo, GP}.
\begin{equation}\tag{{\bf {C}}}\label{C} \mbox{ For any } \xi \in \mathbb R^N\backslash\{0\},\; \mu (\{ v \in \mathscr V, \mbox{ such that } v\cdot\xi \neq 0 \}) > 0.
\end{equation}

For further purposes, it is convenient to introduce the following functional space
$$H=\Big\{(P_1,...,P_M):\mathbb R^N\rightarrow  \mathbb R, \textrm{ such that } \ds\sum_{m=1}^M\ds\int P_m^2\ds\frac{\gamma_m}{\mathscr E_m}\ud v<\infty\Big\}.
$$
Clearly, it defines a Hilbert space, and the components of any elements of $H$ are
integrable functions.
Therefore, it makes sense to consider the following closed subspace 
$$H_0=\Big\{P\in H,   \textrm{ such that }  \ds\sum_{m=1}^M\ds\int P_m\ud v=0\Big\}.
$$
We consider the operators:
\begin{equation}\label{defK}
\mathscr K: (P_1,...,P_M)\longmapsto \Big\{\ds\frac1{\gamma_m}\ds\sum_{\ell=1}^M \ds\int \gamma_{m,\ell}(v,v')P_\ell(v')\ud v',\ m\in \{1,...,M\}\Big\}\end{equation}
and, similarly,
\begin{equation}\label{defQ}
\mathscr Q: (P_1,...,P_M)\longmapsto \Big\{
\ds\frac1{\gamma_m}\ds\sum_{\ell=1}^M \ds\int \gamma_{m,\ell}(v,v')P_\ell(v')\ud v'- P_m(v),\ m\in \{1,...,M\}\Big\}.\end{equation}
We can use shorthand notation: with $P=(P_1,...,P_M)$, we  denote $\mathscr K(P)$ and $\mathscr Q(P)$ 
the vector valued quantities  with components defined above
 and we have
$\mathscr Q_m(P)=\mathscr K_m(P)-P_m=\frac{1}{\gamma_m}Q_m(P)$.
The following statement is an adaptation of \cite[Prop.~1]{DGP}; the detailed proof is given for the sake of completenesss.

\begin{proposition}
\label{p1}
The operators \eqref{defK} and \eqref{defQ} are well defined in $\mathcal L(H)$ and they satisfy the following dissipation property:
denoting $$
B(P,G)=-\ds\sum_{m=1}^M \ds\int Q_m(P)\ds\frac {G_m}{\mathscr E_m}\ud v 
$$
which is continuous on $H\times H$, we have
$$
B(P,P)=\ds\frac12 \ds\sum_{m,\ell=1}^M\ds\iint  \gamma_{m,\ell}(v,v') \mathscr E_\ell (v')
\Big( \ds\frac {P_m}{\mathscr E_m}(v) -  \ds\frac {P_\ell}{\mathscr E_\ell}(v')\Big)^2\ud v'\ud v
\geq \ds\frac12\|\mathscr Q(P)\|_H^2
 \geq 0.$$
\end{proposition}

\noindent
{\bf Proof.}
Let us start with the manipulations that lead to the dissipation property.
With \eqref{defgam} and exchanging the variables, we have
\[\begin{array}{lll}
\ds\sum_{m=1}\ds\int \gamma_m(v) \ds\frac{P_m^2}{\mathscr E_m}(v) \ud v
&=&\ds\sum_{m=1}\ds\int  \left(\ds\sum_{\ell=1}^M\ds\int \gamma_{\ell,m}(v',v) \ud v'\right) \ds\frac{P_m^2}{\mathscr E_m}(v) \ud v
\\
&=&\ds\sum_{\ell,m=1}\ds\iint \gamma_{\ell,m}(v',v)\mathscr E_m(v) \left(\ds\frac{P_m}{\mathscr E_m}(v)\right)^2 \ud v\ud v'
\\
&=
&\ds\sum_{m,\ell=1}\ds\iint \gamma_{m,\ell}(v,v')\mathscr E_\ell(v') \left(\ds\frac{P_\ell}{\mathscr E_\ell}(v')\right)^2 \ud v'\ud v.
\end{array}\]
Moreover, by using  \eqref{A1}, we can also write
\[\begin{array}{lll}
\ds\sum_{m=1}\ds\int \gamma_m(v) \ds\frac{P_m^2}{\mathscr E_m}(v) \ud v
&=&
\ds\sum_{m=1}\ds\int \gamma_m(v)\mathscr E_m(v)\left( \ds\frac{P_m}{\mathscr E_m}(v) \right)^2\ud v
\\
&=&
\ds\sum_{m=1}\ds\int 
\left(\ds\sum_{\ell=1}^M\ds\int 
\gamma_{m,\ell}(v,v')\mathscr E_\ell(v')
\ud v'
\right)
\left( \ds\frac{P_m}{\mathscr E_m}(v) \right)^2\ud v.
\end{array}\]
It follows that $B(P,P) $ can be cast as
\[
B(P,P)=-\ds\sum_{m,\ell=1}\ds\iint \gamma_{m,\ell}(v,v')\mathscr E_\ell(v') 
\left\{
\ds\frac{P_\ell}{\mathscr E_\ell}(v')\ds\frac{P_m}{\mathscr E_m}(v)
-\ds\frac12
\left(\ds\frac{P_\ell}{\mathscr E_\ell}(v')\right)^2
-\ds\frac12
\left(\ds\frac{P_m}{\mathscr E_m}(v)\right)^2
\right\}
 \ud v'\ud v,
 \]
which is the asserted result.

Next, let us detail the functional inequalities.
Still by combining \eqref{defgam} and \eqref{A1}, we observe that
\[
\|P\|^2_H
=\ds\sum_{m,\ell}\ds\iint  \gamma_{m,\ell}(v,v')\mathscr E_\ell(v')\left(\ds\frac{P_m}{\mathscr E_m}(v)\right)^2\ud v'\ud v.
\]
We shall reinterpret
the bilinear  form $B$ by means of the inner product on $H$; namely, we have
\[
B(P,G)=(P|G)_H- (\mathscr K(P)|G).\]
The Cauchy-Schwarz inequality  implies
\[\begin{array}{lll}
|(\mathscr K(P)|G)|&=&\left|
\ds\sum_{m,\ell=1}^M\ds\iint  \gamma_{m,\ell}(v,v')P_\ell(v') \ds\frac{G_m}{\mathscr E_m}(v)
\ud v'\ud v
\right|
\\&=&
\left|
\ds\sum_{m,\ell=1}^M\ds\iint  \sqrt{\gamma_{m,\ell}(v,v')}
\ds\frac{P_\ell}{\sqrt{\mathscr E_\ell}}(v')
\times   \sqrt{\gamma_{m,\ell}(v,v')\mathscr E_\ell(v')}\ds\frac{G_m}{\mathscr E_m}(v)
\ud v'\ud v
\right|
\\
&\leq& \|P\|_H\|G\|_H.
\end{array}\]
These observations prove the continuity of $B$ of $H\times H$ together (by using $G=\mathscr K(P)$)
$\|\mathscr K(P)\|^2_H\leq \|P\|_H \| \mathscr K(P)\|_H$.
Since $\mathscr K(\mathscr E)=\mathscr E$, we conclude that $\|\mathscr K\|_{\mathcal L(H)}=1$. 
Finally, the relation
\[\|\mathscr Q(P)\|_H^2=\|\mathscr K(P)-P\|_H^2=
\|\mathscr K(P)\|_H^2+\|P\|_H^2-2(\mathscr K(P)|P)_H\]
yields
\[B(P,P)=\|P\|_H^2- (\mathscr K(P)|P)_H = 
\ds\frac 12\|\mathscr Q(P)\|_H^2
+\ds\frac12 (\|P\|_H^2-\|\mathscr K(P)\|_H^2)
\geq \ds\frac 12\|\mathscr Q(P)\|_H^2
.\]
\QED

\subsection{Fredholm alternative}

Assumption \eqref{A2} strengthens the dissipation estimate into a coercivity property, which, in turn, allows us to justify the Fredholm alternative.

\begin{coro}\label{cp1}
Assume \eqref{A1}-\eqref{A2}.
For $P\in H$, let $\rho=\sum_{m=1}^M\int P_m(v)\ud v$.
Then, we have 
$B(P,P)\geq \ds\frac{1}{2\mu\kappa}\|P-\rho\mathscr E\|_H^2.$
Moreover, for any $h=(h_1,...,h_M)$ verifying
\[\ds\sum_{m=1}^M\ds\int_{\mathscr V} \ds\frac{|h_m(v)|^2}{\gamma_m(v)\mathscr E_m(v)}\ud v<\infty,\qquad 
\ds\sum_{m=1}^M\ds\int_{\mathscr V} h_m(v)\ud v=0,\]
there exists a unique $P\in H_0$ such that $Q(P)=h$.
\end{coro}

\noindent
{\bf Proof.}
By virtue of \eqref{A1}, we have
\[
\ds\sum_{m=1}^M\ds\int |h_m(v)|\ud v\leq
\left(\ds\sum_{m=1}^M\ds\int\ds\frac{|h_m(v)|^2}{\gamma_m(v)\mathscr E_m(v)}\ud v\right)^{1/2}
\left(\ds\sum_{m=1}^M\ds\int \gamma_m(v)\mathscr E_m(v)\ud v\right)^{1/2}\]
so that the solvability condition makes sense.

By \eqref{defgam}, we already know that $\sum_{m=1}^M\int Q_m(P)\ud v=([\mathscr Q(P)|\mathscr E)_H=0$:
 $\mathrm{Ran}(\mathscr Q)\subset \mathrm{Span}\{\mathscr E\}^\perp$. Conversely, 
 Proposition \ref{p1} shows that $\mathrm{Span}\{\mathscr E\}^\perp\subset \mathrm{Ker}(\mathscr Q)$, and thus 
 $\overline{\mathrm{Ran}(\mathscr Q)}= \mathrm{Span}\{\mathscr E\}^\perp$.
 With \eqref{A2}, we can deduce that $\mathrm{Ran}(\mathscr Q)$ is closed.
 Indeed, with $\rho=\sum_{m=1}^M\int P_m(v)\ud v$, we start by rewriting
 \[
 P_m-\rho \mathscr E_m=\mathscr E_m\ds\sum_{\ell=1}^M\ds\int \Big(
 \ds\frac {P_m}{\mathscr E_m} (v)-\ds\frac {P_\ell}{\mathscr E_\ell} (v')\Big)\mathscr E_\ell(v')
 \ud v'.\]
 Next,  the Cauchy-Schwarz inequality yields
 \[\begin{array}{l}
 \ds\sum_{m=1}^M\ds\int \big|
P_m(v)-\rho \mathscr E_m(v) \big|^2 \ds\frac{\gamma_m(v)}{\mathscr E_m(v)}\ud v
\\
=
 \ds\sum_{m=1}^M \ds\int \gamma_m(v){\mathscr E_m(v)}
\Big|\ds\sum_{\ell=1}^M\ds\int \Big(
 \ds\frac {P_m}{\mathscr E_m} (v)-\ds\frac {P_\ell}{\mathscr E_\ell} (v')\Big)\mathscr E_\ell(v')
 \ud v'\Big|^2\ud v
 \\
 \leq
  \ds\sum_{m=1}^M \ds\int \gamma_m(v)\mathscr E_m(v)
 \left( \ds\sum_{\ell=1}^M\ds\int \Big(
 \ds\frac {P_m}{\mathscr E_m} (v)-\ds\frac {P_\ell}{\mathscr E_\ell} (v')\Big)^2
 \ds\frac{\gamma_\ell(v')\mathscr E_\ell(v')}{1+\gamma_\ell^2(v')}
 \ud v'\right)
 \\
 \hspace*{6cm}\times  \underbrace{\left(
\ds\sum_{\ell=1}^M\ds\int
  \ds\frac{1+\gamma_\ell^2(v')}{\gamma_\ell(v')}\mathscr E_\ell(v')
  \ud v'
  \right)}
 _{=\mu\textrm{ by \eqref{A1}}} 
  \ud v.
 \end{array}
 \]
 Now, we make use of \eqref{A2} to obtain
 \[\begin{array}{lll}
 \|P-\rho\mathscr E\|_H^2
& \leq &
 \mu  \ds\sum_{\ell,m=1}^M \ds\iint
 \ds\frac{\gamma_\ell(v')\gamma_m(v)\mathscr E_m(v)}{1+\gamma_\ell^2(v')}
\mathscr E_\ell(v') \Big(
 \ds\frac {P_m}{\mathscr E_m} (v)-\ds\frac {P_\ell}{\mathscr E_\ell} (v')\Big)^2
 \ud v'\ud v
 \\
 & \leq &
\kappa \mu  \ds\sum_{\ell,m=1}^M \ds\iint
\gamma_{m,\ell}(v')
\mathscr E_\ell(v') \Big(
 \ds\frac {P_m}{\mathscr E_m} (v)-\ds\frac {P_\ell}{\mathscr E_\ell} (v')\Big)^2
 \ud v'\ud v
 \\
 &\leq&
 2\kappa \mu B(P,P),
  \end{array}
 \]
 owing to Proposition~\ref{p1}.
 This proves that $B$ is coercive on $H_0$.
 Then, a standard variational argument justifies the Fredholm alternative.
\QED

\section{Asymptotic analysis}
\label{S4} 

\subsection{A priori estimates}
\begin{proposition}
\label{prop_dissip}
Let $(P^\eps_1,...,P^\eps_m)$ satisfy \eqref{eq1s}, and set $\rho^\eps(t,x)=\sum_{m=1}^M\int P^\eps(t,x,v)\ud v$.
We can find positive constants $\eps_0,C_1,C_2$ such that for any $\eps\in (0,\eps_0)$, we have
\[\ds\frac12\ds\frac{\ud}{\ud t}\ds\sum_{m=1}^M \ds\iint \ds\frac{|P^\eps_m|^2}{\mathscr E_m}\ud v\ud x 
+\ds\frac{C_1}{\eps^2}\ds\sum_{m=1}^M  \ds\iint |P^\eps_m -\rho^\eps \mathscr E_m|^2 \ds\frac{\gamma_m}{\mathscr E_m} \ud v\ud x
\leq C_2 \ds\sum_{m=1}^M \ds\iint \ds\frac{|P^\eps_m|^2}{\mathscr E_m}\ud v\ud x. 
\]
\end{proposition}

\noindent
{\bf Proof.}
The computation adapts  the approach detailed in the scalar case in \cite{DGP}. 
We have
\begin{equation}\label{ent1}
\ds\frac{\ud}{\ud t}\ds\sum_{m=1}^M \ds\iint \ds\frac{|P^\eps_m|^2}{\mathscr E_m}\ud v\ud x=
\ds\frac1\eps\ds\sum_{m=1}^M \ds\iint s_m v\cdot\nabla_x  P^\eps_m \ds\frac{P^\eps_m}{\mathscr E_m}\ud v\ud x
+\ds\frac1{\eps^2} B(P^\eps,P^\eps).
\end{equation}
The last term recasts as  
\[-\ds\frac12\ds\sum_{\ell,m=1}^M \ds\iiint  \gamma_{m,\ell}(v,v') \mathscr E_\ell (v') \Big(
 \ds\frac{P^\eps_\ell(v')}{\mathscr E_\ell(v')}  - \ds\frac{P^\eps_m}{\mathscr E_m}(v)
\Big)^2
\ud v'\ud v\ud x.\]
Let us introduce 
\[\rho^\eps=\ds\sum_{m=1}^M\ds\int P^\eps_m\ud v,\qquad 
G^\eps_m=\ds\frac{P^\eps_m-\rho^\eps \mathscr E_m}{\eps}.\]
Owing to \eqref{A2} we have
\[
\ds\frac1{\eps^2} B(P^\eps,P^\eps)\leq -\ds\frac{1}{2\kappa\mu} \ds\sum_{\ell,m=1}^M \ds\iint  \ds\frac{|G^\eps_m|^2}{\mathscr E_m}(v)
\ud v\ud x.
\]
The first term in the right hand side of \eqref{ent1} can be rewritten as follows
\[\begin{array}{l}\ds\frac1{2\eps}\ds\sum_{m=1}^M \ds\iint s_m v\cdot\nabla_x  \mathscr E_m \ \left(\ds\frac{P^\eps_m}{\mathscr E_m}\right)^2\ud v\ud x
\\
\qquad
=
\ds\frac1{2\eps}\ds\sum_{m=1}^M \ds\iint s_m v\cdot\nabla_x  \mathscr E_m \ \left(
|\rho^\eps|^2 + 2\eps\rho  \ds\frac{G^\eps_m}{\mathscr E_m}
+\eps^2
\left(\ds\frac{G^\eps_m}{\mathscr E_m}\right)^2\right)\ud v\ud x
\\
 \qquad
=
\ds\sum_{m=1}^M \ds\iint s_m \ds\frac{v\cdot\nabla_x  \mathscr E_m}{\mathscr E_m}
\ds\frac{G^\eps_m}{\sqrt{\mathscr E_m}}\rho^\eps\sqrt{\mathscr E_m}\ud v\ud x
+\ds\frac\eps2\ds\sum_{m=1}^M \ds\iint s_m\ds\frac{v\cdot\nabla_x  \mathscr E_m}{\mathscr E_m}
\ds\frac{|G^\eps_m|^2}{\mathscr E_m}\ud v\ud x,
\end{array}\]
where \eqref{B3} has been used to get rid of the stiffest term.
Hence, with \eqref{B1}, we are led to 
\[\begin{array}{l}
\displaystyle \ds\frac1\eps  
\left\vert  
\ds\sum_{m=1}^M\ds\iint s_m v\cdot\nabla_x P^\eps_m \, \ds\frac{P^\eps_m}{\mathscr E_m} \ud v \, \ud x
\right\vert 
\\
\qquad \leq 
\mu_1 
\left(\ds\int \rho^\eps \left( \ds\sum_{m=1}^M\ds\int_{\mathscr V} G^\eps_m \, \gamma_m \ud v \right) \ud x
+
\ds\frac \eps2  \ds\sum_{m=1}^M \ds\iint | G^\eps_m|^2 \, \frac{\gamma_m}{ \mathscr E_m}   \ud v \, \ud x\right)
\end{array}\]
Let us introduce a parameter $\nu>0$, that will be determined later on.
By using \eqref{A1}, this can dominated by 
\[\begin{array}{l}
\ds\frac{\mu_1}{ 4\nu} \, \ds\sum_{m=1}^M \ds\iint |\rho^\eps|^2 \, \gamma_m \ \mathscr E_m \ud v\, \ud x
+\mu_1 \Big(\nu+\ds\frac\eps2\Big) \ds\iint |G^\eps_m|^2 \,\frac{\gamma_m}{ \mathscr E_m} \ud v \, \ud x
\\
\displaystyle \qquad\leq 
\frac{\mu_1 \mu }{ 4\nu}  \ds\sum_{m=1}^M \ds\int |\rho^\eps|^2  \, \ud x 
+\mu_1 \Big(\nu+\ds\frac\eps2\Big)   \ds\sum_{m=1}^M\ds\iint |G^\eps_m|^2 \, \frac{\gamma_m}{ \mathscr E_m} \ud v\, \ud x. 
\end{array}
\]
With the  Cauchy Schwarz inequality and \eqref{A1}, we have 
\begin{eqnarray} \label{rho2}
0\leq \ds\int |\rho^\eps|^2\ud x \leq \ds\sum_{m=1}^M\ds\iint \frac{|P^\eps_m|^2 }{ \mathscr E_m} \ud v\ud x.  
\end{eqnarray}
Thus, we arrive at
\begin{equation} 
\label{e8}
\begin{array}{lll}
\ds\frac1\eps  \left\vert  
 \ds\sum_{m=1}^M\ds\iint  s_mv\cdot\nabla_x P^\eps_m \, \ds\frac{P^\eps_m}{\mathscr E_m}  \ud v \, \ud x
\right\vert  \\
\displaystyle\qquad\leq 
\frac{\mu_1 \mu }{ 4\nu}  \ds\sum_{m=1}^M\ds\iint \frac{|P^\eps_m|^2 }{ \mathscr E_m} \ud v\, \ud x
+\mu_1  \Big(\nu+\ds\frac\eps2\Big)   \ds\sum_{m=1}^M\ds\iint |G^\eps_m|^2 \,\frac{\gamma_m}{ \mathscr E_m} \ud v\, \ud x \, .
\end{array}
\end{equation}
Gathering these information and  coming back to \eqref{ent1}, we get
\[
\begin{array}{l}
\ds\frac12\ds\frac{\ud}{\ud t}  \ds\sum_{m=1}^M\ds\iint \ds\frac{|P^\eps_m|^2(t)}{\mathscr E_m} \ud v\, \ud x
+
\Big(\ds\frac{1}{2\mu\kappa}-\Big(\nu+\ds\frac\eps2\Big)\mu_1\Big)
 \ds\sum_{m=1}^M\ds\iint |G_m^\eps|^2\,\frac{\gamma_m}{ \mathscr E_m} \ud v\, \ud x
\\
\leq
\ds\frac{\mu_1\mu}{4\nu} \ds\sum_{m=1}^M \ds\iint \ds\frac{|P^\eps_m|^2}{\mathscr E_m} \ud v\, \ud x.
\end{array}
\]
This becomes a useful estimate when the coefficient in front of the dissipation term is positive.
To this end, we first choose
 $\nu>0$ so that (for instance) $\frac{1}{2\mu\kappa}-\nu \mu_1
\geq \frac{1}{4\mu\kappa}$. Second, this determines a range  so that
$ \frac{1}{2\mu\kappa}-(\nu+\eps/2)C_1 \geq \frac{1}{8 \mu\kappa}$ for any $0<\eps\leq \eps_0$.
\QED

This statement can be translated into uniform estimates, with a direct application of the Gr\"onwall lemma.

\begin{coro}\label{estim}
Let $P^\eps_{m, \mathrm{ Init}}:\mathbb R^N\times\mathscr V\rightarrow [0,\infty)$ be a sequence of integrable functions 
parametrized by  $\eps\in (0,\eps_0)$
such that
\[
\ds\sup_\eps\ds\sum_{m=1}^M \ds\iint P^\eps_{m, \mathrm{ Init}}\ud v\ud x=M_0<\infty,\qquad
\ds\sup_\eps\ds\sum_{m=1}^M \ds\iint \ds\frac{|P^\eps_{m, \mathrm{ Init}}|^2}{\mathscr E_m}\ud v\ud x=M_1<\infty.\]
Let us expand the  solutions to  \eqref{eq1s} associated to these initial data as 
$P^\eps_m(t,x,v)=\rho^\eps(t,x)\mathscr E_m(x,v)+ \eps G^\eps_m(t,x,v)$, where $\sum_{m=1}^M\int G^\eps_m\ud v=0$. Then, for any $0<T<\infty$, 
\begin{itemize}
\item $\frac{P^\eps_m}{\sqrt {\mathscr E_m}}$ is bounded in $L^\infty(0,T;L^2(\mathbb R^N\times \mathscr V))$,
\item $\sqrt{\frac{\gamma_m}{\mathscr E_m}}g^\eps_m$  is bounded in $L^2((0,T)\times \mathbb R^N\times\mathscr V)$,
\item $\rho^\eps$ is bounded in $L^\infty(0,T;L^1\cap L^2(\mathbb R^N))$,
\item $J^\eps=\frac1\eps\sum_{m=1}^M\int s_m vP^\eps_m\ud v=\sum_{m=1}^M\int s_m v G^\eps_m\ud v$ is bounded in $L^2((0,T)\times\mathbb R^N)$.
\end{itemize}
\end{coro}

We remind the reader that the pair $(\rho^\eps,J^\eps)$ satisfy
\[\partial_t\rho^\eps+\mathrm{div}_xJ^\eps=0.
\]
Then, we wish to pass to the limit in this relation, which amounts to characterize the possible limit of 
\[J^\eps=\sum_{m=1}^M\int s_m vG^\eps_m\ud v.\]
This will be obtained by identifying the limit of the fluctuation in the form
\[G^\eps_m(t,x,v)\xrightarrow[\eps\to 0]{} G(t,x,v)=\chi_m(x,v)\cdot\nabla_x\rho(t,x) +\lambda_m(x,v) \rho(t,x),\]
with $\chi,\lambda$ defined by some auxilliary equations involving the local equilibrium $\mathscr E$.
\\

\subsection{Convergence to the Drift-Diffusion equation}

Equation \eqref{eq1s} holds at least in the sense of distributions, with  
$P^\eps$ belonging
to $C^0([0,+\infty[;L^1(\mathbb  R^N\times \mathscr V))$.
We shall make use of the following weak formulation
\begin{equation}
\label{a1}
\begin{array}{l}
\eps\left(\ds\sum_{m=1}^M\ds\int_D P^\eps_m \phi _m \partial_t \zeta\ud v\ud x\ud t
+
\ds\sum_{m=1}^M\ds\int_D G^\eps_m \phi_m  s_mv\cdot \nabla_x \zeta\ud v\ud x\ud t
\right)
\\
+
\ds\sum_{m=1}^M\ds\int_D \rho^\eps \mathscr E_m \phi_m s_mv\cdot \nabla_x \zeta\ud v\ud x\ud t
+
\ds\sum_{m=1}^M\ds\int_D Q_m(G^\eps)\phi_m\zeta\,  \ud v\ud x\ud t
=0,
\end{array}
\end{equation}
which holds for any $\zeta\in C^\infty_c((0,T)\times\mathbb R^N)$
 $\phi_1,...,\phi_M \in L^\infty(\mathscr V)$, and 
where we have set $D= (0,T)\times\mathbb  R^N\times \mathscr V$,
 $P^\eps_m(t,x,v)=\rho^\eps(t,x)\mathscr E_m(x,v) + \eps \ G^\eps_m(t,x,v)$.

\noindent
{\it Step 1: Weak compactness}

\noindent
Proposition~\ref{prop_dissip} and Corollary~\ref{estim} allow us to assume, possibly at 
the cost of extracting subsequences,
that 
\begin{equation}
\label{c1}
\rho^\eps \rightharpoonup \rho \mbox{ in } L^\infty((0,T);L^2(\mathbb  R^N)) 
\mbox{ weak-$*$},
\end{equation}
\begin{equation}
\label{c2}
J_\eps \rightharpoonup J \mbox{ in } L^2((0,T)\times\mathbb  R^N)\mbox{ weak},
\end{equation}
\begin{equation}
\label{c3}
\sqrt{\ds\frac{\gamma_m}{\mathscr E_m}}G^\eps_m \rightharpoonup \sqrt{\ds\frac{\gamma_m}{\mathscr E_m}} G_m \mbox{ in } L^2(D, \ud v\ud x \ud t)\mbox{ weak}.
\end{equation}
As a matter of fact, it immediately leads to \begin{equation}
\label{a3}
\partial_t \rho
+\nabla_x\cdot J = 0.
\end{equation}
Convergence (\ref{c3}) means that
\begin{equation}
\label{a4}
\ds\lim_{\eps\rightarrow 0}\ds\int_D G^\eps_m \psi_m \ud v\ud x\ud t = 
\ds\int_D G \psi_m \ud v\ud x\ud t \, , 
\end{equation}
provided the test function $\psi=(\psi_1,...,\psi_M)$ satisfies
\begin{equation}
\label{a5}
\ds\sum_{m=1}^M\ds\int_D |\psi_m|^2 \, \frac{\mathscr E_m}{\gamma_m} \ud v\ud x\ud t < \infty.
\end{equation}
In particular, it holds for
\begin{equation}
\label{a6}
\psi_m(t,x,v)=\phi_m(v)\zeta(t,x),
\qquad
\psi_m(t,x,v)=s_mv\phi_m(v)\zeta(t,x),
\end{equation}
for any $\phi_m\in L^\infty(\mathscr V)$, $\zeta\in L^2((0,T)\times\mathbb  R^N)$,
by virtue of \eqref{A1} and \eqref{B2}.

We deduce that  
$$\begin{array}{l}
P^\eps_m \rightharpoonup P_m=\rho \mathscr E_m \textrm{ weakly-$*$ in 
 $L^2(D,\frac{\gamma_m}{\mathscr E_m} \ud v \ud x\ud t)$},
 \\
\rho^\eps\rightharpoonup \rho=\ds\sum_{m=1}^M \ds\int_{\mathscr V} P_m\ud v
 \textrm{ weakly in $ L^2((0,T)\times\mathbb  R^N)$},
\\
J_\eps\rightharpoonup J=\ds\sum_{m=1}^M\ds\int_{\mathscr V} s_m v G_m \ud v.
\end{array}$$
\\

\noindent
{\it Step 2: Passing to the limit in the kinetic equation}

\noindent
Going back to (\ref{a1}) we obtain
\[
- \ds\lim_{\eps\rightarrow 0}
\ds\int_0^T\ds\int_{\mathbb R^N}
\underbrace{
\ds\sum_{m=1}^M\ds\int_{\mathscr V} Q_m(G^\eps)\phi_m\zeta\ud v
}_{B(G^\eps,\phi\zeta F)}
\ud x\ud t
=
\ds\lim_{\eps\rightarrow 0}
\ds\int_0^T\ds\int_{\mathbb R^N}
\rho^\eps \left(
\ds\sum_{m=1}^M\ds\int_{\mathscr V}
s_m v\mathscr E_m \phi_m \ud v\right)\cdot\nabla_x\zeta\ud x\ud t.
\]
Note that by \eqref{A1} and \eqref{B2}, 
the integral with respect to $v$ in the right hand side defines a bounded function.
Therefore  (\ref{c1}) leads to
\begin{equation}
\label{a8}
\ds\sum_{m=1}^M\ds\iiint_D Q_m(G)\phi_m\zeta\ud v\ud x\ud t
+
\ds\sum_{m=1}^M\ds\iiint_D
\rho \  s_m  v\mathscr E_m \phi_m\cdot\nabla_x\zeta\ud v\ud x\ud t=0.
\end{equation}
\\

\noindent
{\it Step 3: Regularity of $\rho$}

\noindent
Since $v\mapsto \frac{v}{|v|}$ lies in $L^\infty(\mathscr V)$, 
we can write
 \begin{equation}
\label{a9}
\ds\sum_{m=1}^M\ds\int_{\mathscr V} Q_m(G)\ds\frac{v}{|v|}\zeta\ud v\ud x\ud t
+
\ds\int_{\mathscr V}
\Theta(x) \rho \nabla_x\zeta\ud x\ud t=0,
\end{equation} 
where $\Theta$ stands for the following (symmetric) matrix
\[
\Theta(x) =\ds\sum_{m=1}^M\ds\int_{\mathscr V} s_m\ds\frac{v\otimes v}{|v|}\mathscr E_m(x,v)\ud v.
\]
By \eqref{A1}, \eqref{B1} and \eqref{B2}, the coefficients of both   $\Theta$ and $D_x \Theta$
belong to  
$L^\infty(\mathbb  R^N)$. We finally appeal to \eqref{C} which  implies that $\Theta(x)$ is definite positive.
By continuity, it follows that for any compact $K\subset \mathbb  R^N$,
we can find $\alpha_K>0$ such that, for all $x\in K$,
\[
\Theta(x)\geq \alpha_K \mathbb I.
\]
Then, (\ref{a9}) can be recast as
\[
\left|
\langle \mathrm{Div}_x(\Theta \rho), \zeta \rangle_{{\mathcal D}', {\mathcal D}}
\right|
=
\left|
\ds\int_0^T\ds\int B\Big(g,\ds\frac{v}{|v|}\mathscr E\zeta\Big)\ud x\ud t
\right|
\leq
\sqrt{M} \Vert \zeta \Vert_{L^2((0,T)\times\mathbb  R^N)}
\Vert \sqrt{\frac{\gamma}{\mathscr E}} G \Vert_{L^2(D,\ud v\ud x\ud t)}, 
\]
by using \eqref{A1} and the continuity of $ B$, and, for a matrix valued function $x\mapsto A(x)$, $\mathrm{Div}_x(A)$ is the shorthand notation 
for the vector with components $\sum_{j=1}^N \partial_{x_j}A_{ij}$.
Accordingly,
$\mathrm{Div}_x(\Theta \rho)=\rho \mathrm{Div}_x(\Theta) + \Theta\nabla_x\rho$ lies in $L^2((0,T)\times\mathbb  R^N)$.
We deduce that 
$\nabla_x\rho\in L^2_{\mathrm{loc}}((0,T)\times\mathbb  R^N)$.
Moreover,
(\ref{a8}) becomes
\[
\ds\sum_{m=1}^M\ds\iiint_D Q_m(G)\phi_m\zeta\ud v\ud x\ud t
-
\ds\sum_{m=1}^M\ds\iiint_D
  \mathrm{div}_x(s_m v\mathscr E_m\rho) \phi_m\zeta\ud v\ud x\ud t=0.
\]
Since this relation holds for all $\phi, \zeta$, we obtain finally the 
following pointwise
relation
\[
Q_m(G)=\mathrm{div}_x(s_m v\rho\mathscr E_m)=s_m v\mathscr E_m\cdot\nabla_x\rho + \rho s_m v\cdot\nabla_x \mathscr E_m.
\]
\\

\noindent
{\it Step 4: Identification of the limit equation}

\noindent
We check that $s_m v\mathscr E_m$ and $s_m v\cdot\nabla_x \mathscr E_m$ define $L^\infty((0,T)\times\mathbb  R^N;
L^2(\mathscr V,\frac{1}{\gamma \mathscr E}\ud v))$ functions by 
Assumptions \eqref{A1}-\eqref{A2}. Hence, \eqref{B1}-\eqref{B2} allow us to 
 apply Corollary~\ref{cp1}
 and 
to define $\chi^{(1)},...,\chi^{(N)}$ and $\lambda$ with values in $H_0$,
solutions of 
$$Q_m(\chi^{(j)})=s_mv^{(j)}\mathscr E_m,\qquad 
Q_m(\lambda)=s_mv\cdot\nabla_x \mathscr E_m.
$$ These functions belong to 
$L^\infty((0,T)\times\mathbb  R^N;H)$.
Furthermore, 
taking $\psi(t,x,v)=\zeta(t,x)$ in (\ref{a4}) gives
\[
\ds\int_0^T\ds\int_{\mathbb  R^N} \zeta\left(\ds\sum_{m=1}^M\ds\int_{\mathscr V} G^\eps_m\ud v\right)\ud x\ud t
=0\rightarrow 
\ds\int_0^T\ds\int_{\mathbb  R^N} \zeta\left(\ds\sum_{m=1}^M\ds\int_{\mathscr V} G_m\ud v\right)\ud x\ud t.
\] 
Going back to  Corollary~\ref{cp1}
we end up with 
\[
G_m(t,x,v)=\ds\sum_{j=1}^N\chi^{(j)}_m(x,v)\partial_{x_j}\rho(t,x) + \lambda_m(x,v)\rho(t,x),
\]
and  we deduce that
\[
J(t,x)=U(x)\rho(t,x) - D(x)\nabla_x \rho(t,x),\]
with 
\[\begin{array}{l}
U(x) = \ds\sum_{m=1}^M\ds\int_{\mathscr V} s_mv \lambda(x,v) \ud v, 
\qquad D(x)=-\ds\sum_{m=1}^M \ds\int_{\mathscr V} s_m v\otimes\chi_m(x,v) \ud v. 
\end{array}
\]
Standard arguments also show that $\rho^\eps$
lies in a compact set of $C^0([0,T];H^{-1}_{\mathrm{loc}}(\mathbb  R^N))$ so that the initial data for the limit problem also makes sense:
it corresponds to 
 the weak limit of $\int_{\mathscr V} f^{\eps}_{\mathrm{Init}} \ud v$.
\\

\noindent
{\it Step 5: Strong convergence}

\noindent
The proof of the strong convergence of $\rho^\eps$
relies on a compensated-compactness argument, see 
 \cite{LiTo, DGP, GP}. This argument avoids the use of the
averaging lemma \cite{GLPS,GPS} which would not apply for discrete velocity models.
Indeed, we have 
\begin{equation}
\label{a12}
\begin{array}{lll}
\Theta(x)\nabla_x\rho^\eps
& = &
-\mathrm{Div}_x(\Theta(x))\rho^\eps +
\ds\sum_{m=1}^M\ds\int_{\mathscr V} \ds\frac{v}{|v|}  \, Q_m(G^\eps) \ud v 
\\
& & -\eps\left(
\partial_t\left[\ds\sum_{m=1}^M\ds\int_{\mathscr V} \ds\frac{v}{|v|} \, P_m^\eps \ud v\right]
+
\mathrm{Div}_x\left[\ds\sum_{m=1}^M\ds\int_{\mathscr V} s_m\ds\frac{v\otimes v}{\vert v\vert} \, G_m^\eps \ud v
\right]\right).
\end{array}
\end{equation}
By using the a priori estimates and Rellich's theorem, we observe 
that the right hand side in (\ref{a12}) lies in a compact set of
$H^{-1}_{\mathrm{loc}}((0,T)\times\mathbb  R^N)$.
The matrix $\Theta$ being  invertible, with the components of  
$D_x\Theta$, and $\Theta^{-1}$  locally bounded,
we deduce that $\nabla_x\rho^\eps$ belongs to a   compact set for the norm of 
$H^{-1}_{\mathrm{loc}}((0,T)\times\mathbb  R^N)$.
Let us introduce the following vector fields (having $N+1$ components) 
\[U^\eps=(\rho^\eps,J^\eps),\qquad V^\eps=(\rho^\eps,0,...,0),\]
which satisfy
\[\begin{array}{l}
\mathrm{div}_{t,x}U_\eps=\partial_t\rho^\eps+\mathrm{div}_x J_\eps=0
\in \mbox{Compact set of} \, H^{-1}_{\mathrm{loc}},
\\
\mathrm{curl}_{t,x} V_\eps=
\left(
\begin{array}{llll}
0 & & -(\nabla_x\rho^\eps )^\intercal 
\\
\nabla_x\rho^\eps &  & 0 &
\end{array}
\right)
\in \mbox{Compact set of} \, (H^{-1}_{\mathrm{loc}})^{(N+1)\times (N+1)}.
\end{array}\]
A direct application of  the div-curl lemma \cite{T,T2}
tells us that 
\[U_\eps\cdot V_\eps=|\rho^\eps|^2\rightarrow 
\left(\begin{array}{l}
\rho \\ J
\end{array}
\right)
\cdot
\left(\begin{array}{l}
\rho \\ 0
\end{array}
\right)
=\rho^2 \;\textrm { in $\mathscr D'((0,T)\times \mathbb  R^N)$.}
\]
It implies 
the strong convergence $\rho^\eps\longrightarrow \rho$ in
$L^2(0,T;L^2_{\mathrm{loc}}(\mathbb  R^N))$.
\QED

\bibliography{biblio}
\bibliographystyle{ieeetr}

\end{document}